\pdfoutput=1
\documentclass[a4paper,american,floatfix,pdftex,superscriptaddress,twoside,%
aip,jcp,%
citeautoscript,
reprint,twocolumn,
]{revtex4-2}%
\usepackage{amsfonts,amsmath,amssymb}
\usepackage{commath}
\usepackage[T1]{fontenc}
\usepackage{lmodern}
\usepackage{graphicx}%
\usepackage[utf8]{inputenc}
\usepackage{microtype}
\usepackage[obeyFinal,textsize=footnotesize]{todonotes}
\usepackage{xspace}
\usepackage{hyperref, hypernat}
\usepackage[todos]{CMImacros}

\graphicspath{{./figures/}} 

\newcommand{\cfeldesy}{\affiliation{Center for Free-Electron Laser Science CFEL,
Deutsches Elektronen-Synchrotron DESY, Notkestr. 85, 22607 Hamburg, Germany}}%
\newcommand{\uhhcui}{\affiliation{Center for Ultrafast Imaging, Universität Hamburg, Luruper
      Chaussee 149, 22761 Hamburg, Germany}}%
\newcommand{\uhhphys}{\affiliation{Department of Physics, Universität Hamburg, Luruper Chaussee 149,
      22761 Hamburg, Germany}}%
\newcommand{\ucl}{\affiliation{Department of Physics and Astronomy, University College London, Gower
      Street, WC1E 6BT London, United Kingdom}}%
\newcommand{\ayemail}{\email[Email: ]{andrey.yachmenev@cfel.de}}%
\newcommand{\cmiweb}{\homepage[URL: ]{https://www.controlled-molecule-imaging.org}}%

\newcommand{\orthopara}{\emph{ortho-para}\xspace}

\begin{document}
\title{The nuclear-spin-forbidden rovibrational transitions of water from first principles}%
\author{Andrey Yachmenev}\ayemail\cmiweb\cfeldesy\uhhcui%
\author{Guang Yang}\cfeldesy\uhhphys%
\author{Emil Zak}\cfeldesy%
\author{Sergei Yurchenko}\ucl%
\author{Jochen Küpper}\cfeldesy\uhhcui\uhhphys%

\date{\today}
\begin{abstract}
   The water molecule occurs in two nuclear-spin isomers that differ by the value of the total
   nuclear spin of the hydrogen atoms, \ie, $I=0$ for \emph{para}-\HHO and $I=1$ for
   \emph{ortho}-\HHO. Spectroscopic transitions between rovibrational states of \emph{ortho} and
   \emph{para} water are extremely weak due to the tiny hyperfine nuclear-spin-rotation interaction
   of only $\ordsim30$~kHz and so far were not observed. We report the first comprehensive
   theoretical investigation of the hyperfine effects and \orthopara transitions in
   H$_2$$^{16}$O due to nuclear-spin-rotation and spin-spin interactions. We also present the
   details of our newly developed general variational approach to the simulation of hyperfine
   effects in polyatomic molecules. Our results for water suggest that the strongest \orthopara
   transitions with room-temperature intensities on the order of
   $10^{-31}$~cm/molecule are about an order of magnitude larger than previously predicted values
   and should be detectable in the mid-infrared $\nu_2$ and near-infrared
   $2\nu_1+\nu_2$ and $\nu_1+\nu_2+\nu_3$ bands by current spectroscopy experiments.
\end{abstract}
\maketitle

\section{Introduction}
The water molecule is abundant in nature. It has two nuclear spin isomers, \emph{ortho}, with a
total nuclear spin of hydrogen atoms $I=1$, and \emph{para}, with a total nuclear spin of hydrogens
$I=0$. In isolated-molecule conditions the \emph{ortho} and \emph{para} nuclear spin isomers show
tremendously long-lasting stability to inter-conversion~\cite{Cacciani:PRA85:012521,
   Miani:JCP120:2732}, can be spatially separated~\cite{Horke:ACIE53:11965,
   Kravchuk:Science331:319}, and exhibit distinct physical and chemical
properties~\cite{Kilaj:NatComm9:2096, Beduz:PNAS109:12894}. Thus the nuclear spin isomers of water
are frequently treated as distinct molecular species.

The concept of stable nuclear spin isomers is appealing to astrophysicists, as it allows to deduce
temperatures, below 50~K, in cometary comae, star- and planet-forming regions from the observations
of relative abundance of \emph{ortho} and \emph{para} species~\cite{Mumma:AA187:419,
   vanDishoeck:WaterFromCloudsToPlanets, Willacy:SpaSciRev197:151, Kawakita:AJ643:1337,
   Putaud:AA632:A8}. Some astronomical observations however reported anomalous
\emph{ortho}-\emph{para} ratios (OPR), corresponding to spin temperatures that are much lower than
gas kinetic temperatures in the same region~\cite{Hogerheijde:Science334:338, Lis:JPCA117:9661,
   Flagey:ApJ762:11, Dishoek:ChemRev113:9043}. These observations pose the intriguing question if
the OPR values could be altered as a result of internal \orthopara conversion, which can possibly be
enhanced by natural factors, such as molecular collisions~\cite{Curl:JCP46:3220,
   Chapovsky:ARPC50:315, Sun:Science310:1938}, interaction with catalytic
surfaces~\cite{Ilisca:ProgSurfSci41:217}, external fields~\cite{Chapovsky:JPhysB33:1001} and
radiation~\cite{Chapovsky:QuantElec49:473}. Low nuclear-spin temperatures have been attributed to
the photodesorption of water from colder icy grains~\cite{Hollenbach:ApJ690:1497}. However, this
theory was benchmarked and disputed in a number of recent laboratory
experiments~\cite{Hama:ApJ738:L15, Hama:Science351:65, Hama:ApJL857:L13, Sliter:JPCA115:9682}.
Arguably there could be another yet unknown mechanism of spin-non-destructive desorption of water
molecules from ice.

The OPR values can change as a result of the interaction between the nuclear spins and an induced
internal magnetic field of the rotating molecule, which is called the hyperfine spin-rotation
interaction. For the main water isotopologue H$_2^{16}$O, considered here, the $^{16}$O has zero
nuclear spin, and the hyperfine coupling between the spins of the protons is very weak, providing a
fundamental rationale for neglecting the \orthopara conversion in practical applications. However,
it can be significantly enhanced by the accidental resonances between the \emph{ortho} and
\emph{para} states, produced by molecular collisions and interactions with strong external fields
and field gradients. The accurate modeling of these processes may unravel previously unknown
mechanisms contributing to the observed anomalous OPR of water in space. Precise knowledge of the
molecular hyperfine states and corresponding transitions is mandatory for the understanding of such
conversion mechanisms. This information can also be important for cold-molecule precision
spectroscopy relying on controlled hyperfine transitions and hyperfine-state changing
collisions~\cite{Liu:FrontPhys16:42300}.

Here, we report a complete linelist of rovibrational hyperfine transitions in
H$_2$$^{16}$O at room-temperature that we computed using an accurate variational
approach~\cite{Yurchenko:JMS245:126, Yachmenev:JCP143:014105, Yurchenko:JCTC13:4368,
   Chubb:JCP149:014101} with an empirically refined potential energy surface
(PES)~\cite{Mizus:PTRSA376:20170149} and a high-level \emph{ab initio} spin-rotation tensor surface.
The spin-spin coupling was modelled as the magnetic dipole-dipole interaction between the two
hydrogen nuclei. We show that the strongest forbidden \orthopara transitions are on the order of
$10^{-31}$~cm/molecule, which is about ten times stronger than previously reported calculations for
the same lines~\cite{Miani:JCP120:2732}. We also present the details of our variational approach for
computing hyperfine effects, which is general and not restricted by the numbers and specific
magnitudes of the molecules' nuclear spins.

\section{Theoretical details}
\label{sec:theory}
\subsection{Spin-rotation and spin-spin coupling}
In this section we describe the implementation of the hyperfine spin-rotation and spin-spin coupling
terms within the general variational framework of the nuclear motion approach
TROVE~\cite{Yurchenko:JMS245:126, Yachmenev:JCP143:014105, Yurchenko:JCTC13:4368,
   Chubb:JCP149:014101}. Implementation details of the hyperfine nuclear quadrupole coupling can be
found in our previous works~\cite{Yachmenev:JCP147:141101, Yachmenev:JCP151:244118}.

The spin-rotation coupling is the interaction between the rotational angular momentum $\mathbf{J}$
of the molecule and the nuclear spins $\mathbf{I}_n$ of different
nuclei~\cite{Flygare:ChemRev74:653}
\begin{align}
  \label{hsr_cart}
  H_\text{sr} = \sum_{n}^{N_I} \mathbf{I}_n \cdot \mathbf{M}_n \cdot \mathbf{J},
\end{align}
where $\mathbf{M}_n$ is the second-rank spin-rotation tensor relative to the nucleus $n$ and the sum
runs over all nuclei $N_I$ with non-zero spin. The interaction between the nuclear spins
$\mathbf{I}_n$ of different nuclei is given by the spin-spin coupling as
\begin{align}
  \label{hss_cart}
  H_\text{ss} = \sum_{n>n'}^{N_I} \mathbf{I}_n \cdot \mathbf{D}_{n,n'} \cdot \mathbf{I}_{n'},
\end{align}
where $\mathbf{D}_{n,n'}$ is the second-rank spin-spin tensor, which is traceless and symmetric.
Using the spherical-tensor representation~\cite{Zare:AngularMomentum}, the spin-rotation and
spin-spin Hamiltonians can be expressed as
\begin{align}
  \label{hsr_spher}
  H_\text{sr} = &\frac{1}{2}\sum_{n}^{N_I} \sum_{\omega=0}^2 \sqrt{2\omega+1}
                  \left(-\frac{1}{\sqrt{3}}\right) \mathbf{I}_n^{(1)}\cdot \\
  \nonumber
                &\cdot\left( (-1)^\omega \left[ \mathbf{M}_n^{(\omega)} \otimes \mathbf{J}^{(1)} \right]^{(1)} +
                  \left[ \mathbf{J}^{(1)} \otimes \mathbf{M}_n^{(\omega)} \right]^{(1)} \right)
\end{align}
and
\begin{equation}
  \label{hss_spher}
  H_\text{ss} = \sum_{n>n'}^{N_I} \mathbf{D}_{n,n'}^{(2)} \cdot \left[ \mathbf{I}_n^{(1)}
     \otimes \mathbf{I}_{n'}^{(1)} \right]^{(2)},
\end{equation}
where $\mathbf{M}_n^{(\omega)}$, $\mathbf{D}_{n,n'}^{(2)}$, $\mathbf{J}^{(1)}$, and
$\mathbf{I}_n^{(1)}$ denote the spherical-tensor representations of operators in \eqref{hsr_cart}
and \eqref{hss_cart} and the square brackets are used to indicate the tensor product of two
spherical-tensor operators. Because the spin-rotation tensor is generally not symmetric, the second
term in the sum \eqref{hsr_spher} is added to ensure that the Hamiltonian is Hermitian.

The nuclear-spin operator $\mathbf{I}_n$ and the rotational-angular-momentum operator $\mathbf{J}$
are coupled using a \emph{nearly-equal} coupling scheme, \ie,
$\mathbf{I}_{1,2}=\mathbf{I}_1+\mathbf{I}_2$, $\mathbf{I}_{1,3}=\mathbf{I}_{1,2}+\mathbf{I}_3$,
\ldots, $\mathbf{I}\equiv\mathbf{I}_{1,N}=\mathbf{I}_{1,N-1}+\mathbf{I}_{N}$, and
$\mathbf{F}=\mathbf{J}+\mathbf{I}$. The nuclear-spin functions $\ket{I,m_I,\mathcal{I}}$ depend on
the quantum numbers $I$ and $m_I$ of the collective nuclear-spin operator $\mathbf{I}$ and its
projection onto the laboratory $Z$ axis, respectively. The set of auxiliary quantum numbers
$\mathcal{I}=\{I_1,I_{1,2},I_{1,3},\ldots,I_{1,N-1}\}$ for the intermediate spin angular momentum
operators provide a unique assignment of each nuclear-spin state. The total spin-rovibrational wave
functions $\ket{F,m_F,u}$ are built as symmetry-adapted linear combinations of the coupled products
of the rovibrational wave functions $\ket{J,m_J,l}$ and the nuclear-spin functions
$\ket{I,m_I,\mathcal{I}}$. Here, $J$ and $F$ are the quantum numbers of $\mathbf{J}$ and
$\mathbf{F}$ operators with $m_J$ and $m_F$ of their $Z$-axis projections. $l$ and $u$ denote the
rovibrational and hyperfine state indices, respectively, and embrace all quantum numbers, \eg,
rotational $k$ and vibrational quantum numbers $v_1,v_2,\ldots$, that are necessary to characterize
a nuclear spin-rovibrational state.

The symmetrization postulate requires the total wavefunction of the \HHO molecule to change sign
upon exchange of the protons, \ie, to transform as one of the irreducible representations $B_1,B_2$
of its $\textbf{C}_\text{2v}$(M) symmetry group. Accordingly, the \emph{ortho} spin state
$\ket{I=1}$ of $A_1$ symmetry can be coupled with the rovibrational states of $B_1$ and $B_2$
symmetries and the \emph{para} state $\ket{I=0}$ of $B_2$ symmetry can be coupled with the
rovibrational states of $A_1$ and $A_2$ symmetries.

The matrix representations of the spin-rotation and spin-spin Hamiltonians in the basis of the
$\ket{F,m_F,u}$ functions are diagonal in $F$ and $m_F$, with the explicit expressions given by
\begin{align}\label{hsr_me}
  \langle &F,m_F,u'|H_\text{sr}|F,m_F,u\rangle = \\ \nonumber
          &=\frac{1}{2}(-1)^{I+F}\sqrt{(2J+1)(2J'+1)} \left\{ \begin{array}{ccc}I'&J'&F\\J&I&1\end{array}
                                                                                              \right\} \\ \nonumber
          &\times\sum_{n}^{N_I}\sum_{\omega=0}^2 N_\omega
            \left[
            (-1)^\omega J \left\{ \begin{array}{ccc}\omega&1&1\\J&J'&J\end{array}
                                                                      \right\}\left( \begin{array}{ccc}J&1&J\\-J&0&J\end{array} \right)^{-1}
                                                                                                                    \right. \\ \nonumber
          & +\left.
            J' \left\{ \begin{array}{ccc}1&\omega&1\\J&J'&J'\end{array}\right\}\left(
                                                           \begin{array}{ccc}J'&1&J'\\-J'&0&J'\end{array} \right)^{-1}
                                                                                             \right] \\ \nonumber
          &\times\mathcal{M}_{\omega,n}^{(J'l',Jl)} \langle I'||\mathbf{I}_n^{(1)}||I\rangle
\end{align}
and
\begin{align}\label{hss_me}
  \langle &F,m_F,u'|H_\text{ss}|F,m_F,u\rangle =  \\ \nonumber
          &=(-1)^{I+J'+J+F}\sqrt{(2J+1)(2J'+1)} \left\{ \begin{array}{ccc}I'&J'&F\\J&I&2\end{array}
                                                                                        \right\} \\ \nonumber
          &\times\sum_{n>n'}^{N_I}\mathcal{D}_{n,n'}^{(J'l',Jl)}
            \langle I'|| [ \mathbf{I}_n^{(1)} \otimes \mathbf{I}_{n'}^{(1)} ]^{(2)} ||I\rangle,
\end{align}
with the normalization constant $N_\omega=1$, $-\sqrt{3}$, and $\sqrt{5}$ for $\omega=0$, 1, and 2,
respectively. The expressions for the reduced matrix elements of the nuclear-spin operators
$\langle I'||\mathbf{I}_n^{(1)}||I\rangle$ and
$\langle I'|| [ \mathbf{I}_n^{(1)} \otimes \mathbf{I}_{n'}^{(1)} ]^{(2)} ||I\rangle$ depend on the
total number of coupled spins and can be computed using a general recursive procedure as described,
for example, in ref.~\onlinecite{Yachmenev:JCP147:141101}. Here, for the two equivalent hydrogen
spins $I_1=I_2=1/2$, the reduced matrix elements are
\begin{align}
  & \langle I'||\mathbf{I}_n^{(1)}||I\rangle = (-1)^{I\delta_{n,1}+I'\delta_{n,2}} I_1  \\
  \nonumber
  & \times \sqrt{(2I+1)(2I'+1)}
    \left\{ \begin{array}{ccc}I_1&I'&I_1\\I&I_1&1\end{array} \right\}
                                                 \left( \begin{array}{ccc}I_1&1&I_1\\-I_1&0&I_1\end{array} \right)^{-1},
\end{align}
with the explicit values $\langle 0||\mathbf{I}_n^{(1)}||0\rangle=0$,
$\langle 1||\mathbf{I}_n^{(1)}||1\rangle=\sqrt{3/2}$,
$\langle 0||\mathbf{I}_n^{(1)}||1\rangle=\pm\sqrt{3}/2$ for $n=1$ and $2$, respectively, and
$\langle 1||\mathbf{I}_n^{(1)}||0\rangle=\mp\sqrt{3}/2$.

The expressions for the $\mathcal{M}_{\omega,n}^{(J'l',Jl)}$ and $\mathcal{D}_{n,n'}^{(J'l',Jl)}$
tensors in Eqs.~\eqref{hsr_me} and \eqref{hss_me} depend on the chosen rovibrational wave functions
$\ket{J,m_J,l}$, which are represented by the molecular rovibrational eigenfunctions calculated with
the variational approach TROVE. The functions $\ket{J,m_J,l}$ are linear combinations of products of
vibrational wave functions $\ket{\nu}=\ket{v_1,v_2,\ldots,v_{M}}$ ($M$ is the number of vibrational
modes) and symmetric-top rotational functions
\begin{align}
  |J,m_J,l\rangle = \sum_{\nu,k} c_{\nu,k}^{(J,l)}
  \ket{\nu}\ket{J,k,m_J}.
\end{align}
In this basis, the $\mathcal{M}_{\omega,n}^{(J'l',Jl)}$ and $\mathcal{D}_{n,n'}^{(J'l',Jl)}$ tensors are
\begin{align}\label{m_tens}
  & \mathcal{M}_{\omega,n}^{(J'l',Jl)} = \sum_{\nu' k'}\sum_{\nu k}
    \left[c_{\nu' k'}^{(J',l')}\right]^* \, c_{\nu k}^{(J,l)} \, (-1)^{k'} \\ \nonumber
  & \times \sum_{\sigma=-\omega}^\omega\sum_{\alpha,\beta=x,y,z}
    \left(\begin{array}{ccc} J & \omega & J' \\ k & \sigma & -k'\end{array}\right)
                                                             U_{\omega\sigma,\alpha\beta}^{(2)} \langle \nu'|\bar{M}_{\alpha\beta,n}|\nu\rangle
\end{align}
and
\begin{align}\label{d_tens}
  & \mathcal{D}_{n,n'}^{(J'l',Jl)} = \sum_{\nu' k'}\sum_{\nu k}
    \left[c_{\nu' k'}^{(J',l')}\right]^* \, c_{\nu k}^{(J,l)} \, (-1)^{k'} \\ \nonumber
  & \times \sum_{\sigma=-2}^2\sum_{\alpha,\beta=x,y,z}
    \left(\begin{array}{ccc} J & 2 & J' \\ k & \sigma & -k'\end{array}\right)
                                                        U_{2\sigma,\alpha\beta}^{(2)} \langle \nu'|\bar{D}_{\alpha\beta,nn'}|\nu\rangle
\end{align}
where $\bar{M}_{\alpha\beta,n}$ and $\bar{D}_{\alpha\beta,nn'}$ ($\alpha,\beta=x,y,z$) are
spin-rotation and spin-spin interaction tensors in the molecule-fixed frame and the $9\times9$
constant matrix $U_{\omega\sigma,\alpha\beta}^{(2)}$ ($\omega=0,\dots,2$,
$\sigma=-\omega,\ldots,\omega$) defines the transformation of a general second-rank
Cartesian tensor operator into its spherical-tensor representation, see,
\eg, (5.41)--(5.44) in ref.~\onlinecite{Zare:AngularMomentum}.

The total Hamiltonian $H$ is composed of a sum of the pure rovibrational Hamiltonian $H_\text{rv}$
and hyperfine terms $H_\text{sr}$ and $H_\text{ss}$. In the basis of TROVE wave functions, the
rovibrational Hamiltonian $H_\text{rv}$ is diagonal, its elements are given by the rovibrational
energies
\begin{multline}
  \label{tot_ham}
  \langle F,m_F,u'|H|F,m_F,u\rangle \\
  = E_u\delta_{u,u'} + \langle F,m_F,u'|H_\text{sr}|F,m_F,u\rangle \\
  + \langle F,m_F,u'|H_\text{ss}|F,m_F,u\rangle,
\end{multline}
where $\delta_{u,u'}=\delta_{J,J'}\delta_{l,l'}\delta_{I,I'}\delta_{\mathcal{I},\mathcal{I'}}$.

The above equations were implemented in the \texttt{hyfor} module of the Python software package
Richmol~\cite{Owens:JCP148:124102, richmol2021github}, which uses rovibrational molecular states
calculated in TROVE as a variational basis. Alternative approaches using Watson-type effective
Hamiltonians~\cite{Watson:VibSpecStruct6:1} are also implemented in the Richmol package.

The hyperfine energies and wave functions are computed in a three step procedure. First, we solve
the full rovibrational problem using TROVE and obtain the rovibrational energies and wave functions
for all states with energies below a selected threshold. In the next step, the rovibrational matrix
elements of the spin-rotation and spin-spin tensors are computed in the form given by
Eqs.~\eqref{m_tens} and \eqref{d_tens}. These matrix elements are later used to build the
spin-rotation and spin-spin interaction Hamiltonians using Eqs.~\eqref{hsr_me} and \eqref{hss_me}.
The total Hamiltonian is composed of the sum of a purely rovibrational part, which is diagonal and
given by the rovibrational state energies, and non-diagonal spin-rotation and spin-spin parts. In
the final step, the hyperfine energies and wave functions are obtained by diagonalizing the total
Hamiltonian.

The computation of the dipole transition intensities also proceeds in two steps. First, the
rovibrational matrix elements of the dipole moment surface are computed and cast into a tensor form
similar to \eqref{d_tens},
\begin{align}\label{k_tens}
  & \mathcal{K}_{\omega}^{(J'l',Jl)} = \sum_{\nu' k'}\sum_{\nu k}
    \left[c_{\nu' k'}^{(J',l')}\right]^* \, c_{\nu k}^{(J,l)} \, (-1)^{k'} \\ \nonumber
  & \times \sum_{\sigma=-\omega}^\omega\sum_{\alpha,\beta=x,y,z}
    \left(\begin{array}{ccc} J & \omega & J' \\ k & \sigma & -k'\end{array}\right)
                                                             U_{\omega\sigma,\alpha}^{(1)} \langle \nu'|\bar{\mu}_{\alpha}|\nu\rangle,
\end{align}
where $\bar{\mu}_{\alpha}$ ($\alpha=x,y,z$) is the permanent dipole moment in the molecule-fixed
frame and the $3\times3$ constant matrix $U_{\omega\sigma,\alpha}^{(1)}$ ($\omega=1$,
$\sigma=-\omega,\ldots,\omega$) defines the transformation of a general first-rank Cartesian tensor
operator into its spherical-tensor representation, see, \eg,
(5.4) in ref.~\onlinecite{Zare:AngularMomentum}. In the second step, the dipole matrix elements are
transformed into the basis of hyperfine wave functions, \ie,
\begin{multline} {\mathcal{K}_{\omega}^{(F',u',F,u)} = \sum_{I',{\mathcal
            I}',J',l'}\sum_{I,{\mathcal I},J,l} \left[ c_{I',{\mathcal
               I}',J',l'}^{(F',u')} \right]^* c_{I,\mathcal{I},J,l}^{(F,u)} (-1)^{I}} \\
   {\times\sqrt{(2J'+1)(2J+1)}\left\{\begin{array}{ccc}J'&F'&I \\ F&J&\omega\end{array}\right\}
      \mathcal{K}_{\omega}^{(J',l',J,l)}\delta_{I',I}\delta_{\mathcal{I}',\mathcal{I}}},
\end{multline}
where $c_{I,\mathcal{I},J,l}^{(F,u)}$ are hyperfine wave function coefficients obtained by
diagonalization of the total Hamiltonian. Finally, the line strengths for transitions between
hyperfine states $\ket{f}=\ket{F',u'}$ and $\ket{i}=\ket{F,u}$ are computed
as~\cite{Yachmenev:JCP151:244118}
\begin{align}
  S(f\leftarrow i) = (2F'+1)(2F+1)\left|\mathcal{K}_{1}^{(F'u',Fu)} \right|^2,
\end{align}
where we sum over all degenerate $m_F$ and $m_F'$ components. The expression for the integrated
absorption coefficient of the dipole transition in units of cm/molecule reads
\begin{align}
  I(f\leftarrow i) = \frac{8\pi^3 \nu_{if}
  e^{-hcE_{i}/kT}\left(1-e^{-hc\nu_{if}/kT}\right)}{3hcZ(T)}S(f\leftarrow i),
  \label{eq:intensity}
\end{align}
where $\nu_{if}=|E_i - E_f|$ is the transition wavenumber, $E_i$ and $E_f$ are energy term values of
the initial and final states in \invcm, $Z(T)$ is the temperature dependent partition function, $h$
(erg$\cdot$s) is the Planck constant, $c$ (cm/s) is the speed of light and $k$ (erg/K)
is the Boltzmann constant.

\subsection{Electronic structure calculations}
The molecule-fixed frame spin-rotation tensors $\bar{M}_{\alpha\beta,n}$ ($\alpha,\beta=x,y,z$,
$n=1,2$) were calculated \emph{ab initio} on a grid of 2000 different molecular geometries with
electronic energies ranging up to 30\,000~cm$^{-1}$ above the equilibrium energy. We used the
all-electron CCSD(T) (coupled-cluster singles, doubles, and perturbative triples) method with the
augmented core-valence correlation-consistent basis set aug-cc-pwCVTZ~\cite{Peterson:JCP117:10548}
and aug-cc-pVTZ~\cite{Dunning:JCP90:1007, Kendall:JCP96:6796} for the oxygen and hydrogen atoms,
respectively. The basis sets were downloaded from the Basis Set Exchange
library~\cite{Pritchard:JCIM59:4814, Feller:JCompChem17:1571, Schuchardt:JCIM47:1045}. The
calculations employed second-order analytical derivatives~\cite{Scuseria:JCP94:442} together with
the rotational London orbitals~\cite{Gauss:JCP105:2804, Gauss:MolPhys91:449}, as implemented in the
quantum chemistry package CFOUR~\cite{CFOUR:2020}.

The electronic structure calculations used the principal axes of inertia coordinate frame. For
variational calculations another frame was employed, defined such that the $x$ axis is parallel to
the bisector of the valence bond angle with the molecule lying in the $xz$ plane at all instantaneous
molecular geometries. In this frame, the $z$ axis coincides with the molecular axis at the linear
geometry. The computed spin-rotation tensors were rotated from the principal axis of inertia to the
new frame. The permutation symmetry is such, that exchange of the two hydrogen atoms transforms
$\bar{M}_{\alpha\beta,1}$ into $\bar{M}_{\alpha\beta,2}$ followed by a sign change for non-diagonal
elements ($\alpha\neq\beta$).

The expression for the spin-rotation tensor, as computed in CFOUR, contains multiplication by the
inverse of the tensor of inertia, see (3) and (7) in ref.~\onlinecite{Gauss:MolPhys91:449}. For
linear and closely linear geometries of the molecule, the inertial tensor becomes singular, which
creates a discontinuity in the dependence of $xz$ and $zz$ elements of spin-rotation tensor on the
bending angle. To circumvent this problem, we have multiplied the computed spin-rotation tensors on
the right side by the corresponding inertial tensors. The resulting data for the inertia-scaled
spin-rotation tensor was parameterized through least-squares fitting, using a power series
expansions to fourth order in terms of valence bond coordinates, with $\sigma_\text{rms}\leq0.3$~kHz
for all tensor components. Later, when computing the rovibrational matrix elements of the
spin-rotation tensor, we have multiplied the inertia-scaled tensor with the inverse moment of
inertia. The divergence of the spin-rotation tensor in the vicinity of linear geometries is exactly
canceled by the basis functions chosen to satisfy the kinetic cusp condition at the linear
geometry~\cite{Chubb:JCP149:014101, Yurchenko:JCP153:154106}.

The spin-spin tensor elements were computed as magnetic dipole-dipole interaction between two
hydrogen nuclei H$_1$ and H$_2$,
\begin{align}
  D_{\alpha\beta,12} = \frac{\mu_0}{4\pi}\frac{\mu_1\mu_2}{I_1 I_2
  r_{12}^3}\left(\mathbf{I}-3\mathbf{n} \otimes \mathbf{n}\right)_{\alpha\beta},
\end{align}
where $\mu_1=\mu_2=2.79284734$ are the magnetic dipole moments of H$_1$ and H$_2$ in units of the
nuclear magneton, $I_1=I_2=1/2$ are the corresponding hydrogen nuclear spins, $r_{12}$ is the
distance between the hydrogen nuclei, and $\mathbf{n}$ is the unit vector directed from one hydrogen
to another. The indirect spin-spin coupling constants mediated by the electronic motions were not
considered here, as they are typically two orders of magnitude smaller than the direct
constants~\cite{Yachmenev:JCP132:114305}.

\subsection{Nuclear motion calculations}
We employed TROVE to calculate the rovibrational states using the exact kinetic-energy operator
formalism recently developed for triatomic molecules~\cite{Yurchenko:JCP153:154106}. This
formalism is based on the use of associated Laguerre polynomials $L_{n}^{l}(x)$ as bending basis
functions, which ensures a correct behavior of the rovibrational wave functions at linear molecular
geometry~\cite{Yurchenko:JCP153:154106}. The bisecting frame embedding was selected as a non-rigid
reference frame, with the $x$ axis oriented parallel to the bisector of the valence bond angle and
the molecule placed in the $xz$ plane. In this frame, the $z$ axis coincides with the linearity axis
at linear molecular geometry. Accurate empirically refined PES of H$_2^{16}$O was
employed~\cite{Mizus:PTRSA376:20170149}.

The primitive-stretching vibrational basis functions were generated by numerically solving the
corresponding one-dimensional Schr\"{o}dinger equations on a grid of 2000 points using the
Numerov-Cooley approach~\cite{Noumerov:MNRAS84:592,Cooley:MathComp15:363}. The primitive basis
functions were then symmetry-adapted to the irreducible representations of the
$\textbf{C}_\text{2v}$(M) molecular symmetry group using an automated numerical procedure
\cite{Yurchenko:JCTC13:4368}. The total vibrational basis set was formed as a direct product of the
symmetry-adapted stretching and bending basis functions, contracted to include states up to a polyad
48. It was used to solve the $J=0$ eigenvalue problem for the complete vibrational Hamiltonian of
H$_2$O. A product of the $J=0$ eigenfunctions and symmetry-adapted rigid rotor wavefunctions formed
the final rovibrational basis set. The rovibrational wavefunctions $\ket{J, m_J, l}$ for rotational
excitations up to $J=40$ and four irreducible representations $A_1$, $A_2$, $B_1$ and $B_2$ were
computed by diagonalizing the matrix representation of the total rovibrational Hamiltonian
$H_\text{rv}$ in the rovibrational basis set. More details about the variational approach and the
basis-symmetrization procedure for the case of triatomic molecules can be found in
ref.~\onlinecite{Yurchenko:JCP153:154106}.

\begin{table*}
   \caption{Strongest predicted \orthopara transitions in H$_2$$^{16}$O at
      $T=296$~K with the $10^{-31}$ cm/molecule intensity cut-off. }
   \label{tab1}
   \tabcolsep=3pt \renewcommand{\arraystretch}{1}
   \begin{tabular}{ccccccccr c ccccccccr rr}
     \hline
	$\nu_1'$ & $\nu_2'$ & $\nu_3'$ & $F'$ & $J'$ & $k_a'$ & $k_c'$ & $I'$ & $E'$ (\invcm) &&
	$\nu_1$ & $\nu_2$ & $\nu_3$ & $F$ & $J$ & $k_a$ & $k_c$ & $I$ & $E$
	(\invcm) &
	Freq. (\invcm) & Int. (cm/molec.) \\
	\hline
	\hline
	0 &   1 &   0 &   3 &   4 &  2 &   3 & $o$ &  1908.016319 &&  0 &   0 &   0 &   4 &   4 &
	4 &   0
	& $p$ &   488.134170 &  1419.882149 & $2.26\times 10^{-31}$  \\
	0 &   1 &   0 &   3 &   3 &  3 &   1 & $p$ &  1907.450231 &&  0 &   0 &   0 &   3 &   4 &
	3 &   2
	&
	$o$ &   382.516901 &  1524.933330 & $1.36\times 10^{-31}$  \\
	0 &   1 &   0 &   3 &   3 &  3 &   1 & $p$ &  1907.450231 &&  0 &   0 &   0 &   3 &   4 &
	1 &   4
	&
	$o$ &   224.838381 &  1682.611850 & $1.12\times 10^{-31}$  \\
	0 &   1 &   0 &   3 &   4 &  2 &   3 & $o$ &  1908.016319 &&  0 &   0 &   0 &   3 &   3 &
	2 &   2
	&
	$p$ &   206.301430 &  1701.714889 & $1.02\times 10^{-31}$  \\
	0 &   1 &   0 &   3 &   3 &  3 &   1 & $p$ &  1907.450231 &&  0 &   0 &   0 &   2 &   3 &
	1 &   2
	&
	$o$ &   173.365811 &  1734.084420 & $2.05\times 10^{-31}$  \\
	0 &   1 &   0 &   3 &   4 &  2 &   3 & $o$ &  1908.016319 &&  0 &   0 &   0 &   2 &   2 &
	2 &   0
	&
	$p$ &   136.163927 &  1771.852392 & $3.28\times 10^{-31}$  \\
	2 &   1 &   0 &   3 &   4 &  1 &   4 & $o$ &  8979.657423 &&  0 &   0 &   0 &   4 &   4 &
	1 &   3
	&
	$p$ &   275.497051 &  8704.160372 & $3.36\times 10^{-31}$  \\
	2 &   1 &   0 &   3 &   4 &  1 &   4 & $o$ &  8979.657423 &&  0 &   0 &   0 &   3 &   3 &
	1 &   3
	&
	$p$ &   142.278493 &  8837.378930 & $1.01\times 10^{-31}$  \\
	2 &   1 &   0 &   3 &   4 &  1 &   4 & $o$ &  8979.657423 &&  0 &   0 &   0 &   2 &   2 &
	1 &   1
	&
	$p$ &    95.175936 &  8884.481487 & $6.41\times 10^{-31}$  \\
	1 &   1 &   1 &  15 &  14 &  3 &  11 & $o$ & 11067.083574 &&  0 &   0 &   0 &  14 &  14 &
	0 &  14
	&
	$p$ &  2073.514207 &  8993.569367 & $1.92\times 10^{-31}$  \\
	1 &   1 &   1 &  15 &  15 &  2 &  13 & $p$ & 11067.089122 &&  0 &   0 &   0 &  14 &  13 &
	1 &  12
	&
	$o$ &  2042.309821 &  9024.779300 & $2.04\times 10^{-31}$  \\
     \hline
   \end{tabular}
\end{table*}

\subsection{Linelist simulations}
The linelist of hyperfine rovibrational transitions for H$_2$$^{16}$O was computed with an energy
cutoff at $15\,000$~\invcm and includes transitions up to $F=39$ ($J=40$). To further improve the
accuracy of the linelist, after solving the pure rovibrational problem and before entering the
hyperfine calculations, the rovibrational energies $E_u$ in \eqref{tot_ham} were replaced with the
high-resolution experimental IUPAC values from ref.~\onlinecite{Tennyson:JQSRT117:29}, where
available. Such empirical adjustment of the rovibrational energies have been adopted and tested,
\eg, for the production of molecular linelists as part of the ExoMol
project~\cite{Tennyson:JQSRT255:107228}. Recently, this approach was proven accurate for computing
the ultra-weak quadrupole transitions in water~\cite{Campargue:PCCP22:12476, Campargue:PRR2:023091}
and carbon dioxide~\cite{Fleurbaey:JQSRT266:107558,Yachmenev:JCP154:211104}, which enabled their
first laboratory (H$_2$O and CO$_2$) and astrophysical (CO$_2$) detection.

The final linelist has been calculated at room temperature
($T=296$~K) with the corresponding partition function
$Z=174.5813$~\cite{Polyansky:MNRAS480:2597}, and a threshold of
$10^{-36}$~cm/molecule for the absorption intensity based on \eqref{eq:intensity}. The linelist
stored in the ExoMol~\cite{Tennyson:JMS327:73}
format is provided in the supplementary information.

\section{Results and discussion}
\begin{figure}
   \includegraphics[width=\linewidth]{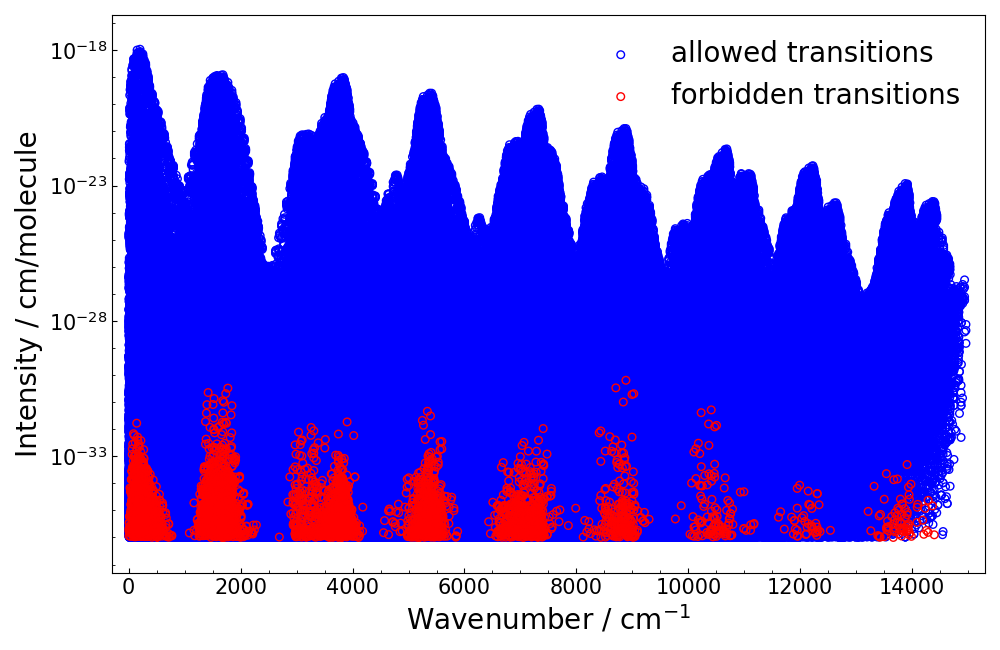}
   \caption{Overview of the H$_2^{16}$O dipole absorption spectrum at $T=296$~K. The
      \emph{ortho-ortho} and \emph{para-para} transitions are marked with blue circles, whereas the
      \orthopara transitions are given by red circles.}
   \label{fig1}
\end{figure}
\begin{figure*}
   \includegraphics[width=0.8\linewidth]{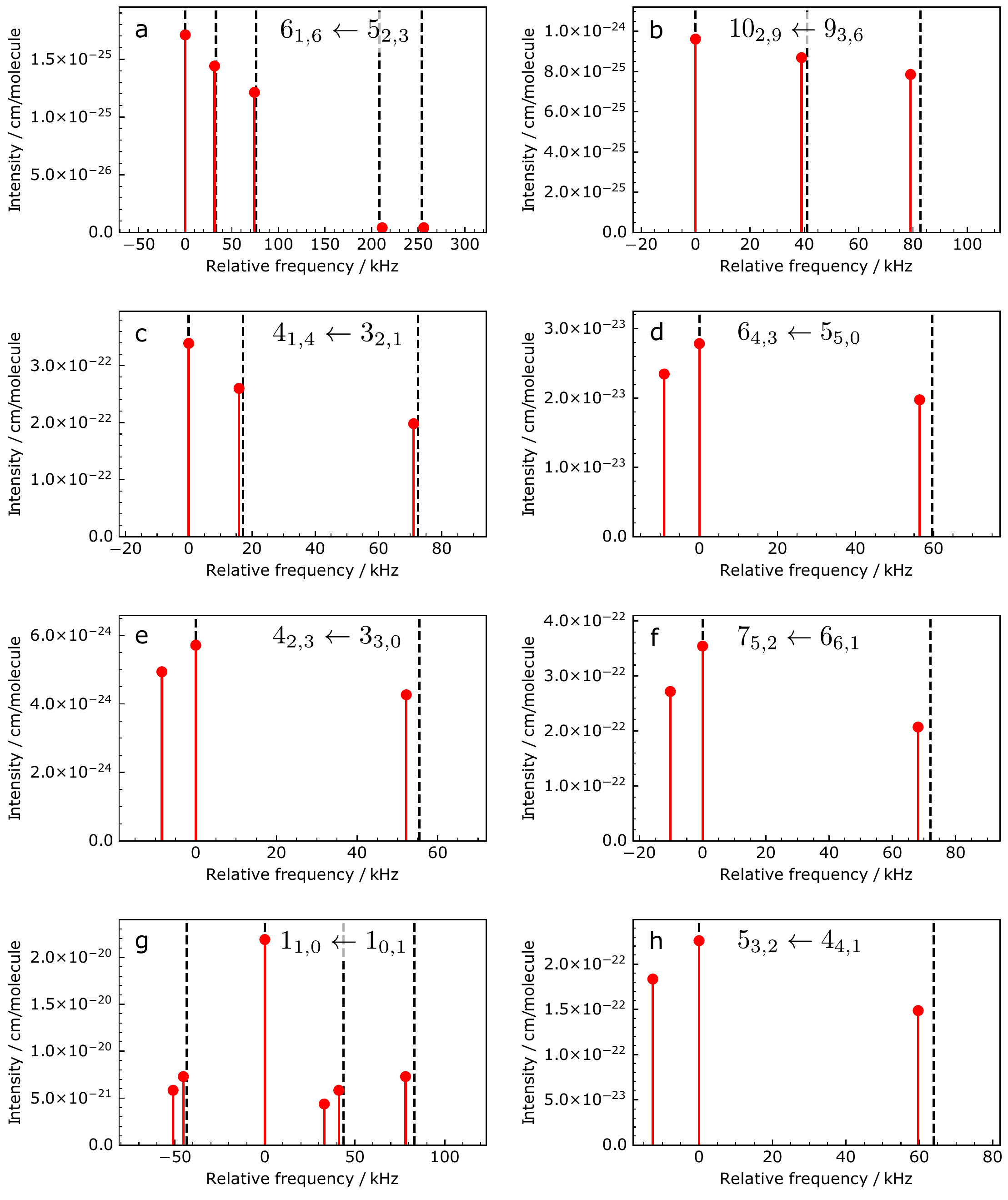}%
   \caption{Comparison of calculated hyperfine transitions (red stems) with experimental data
      (dashed lines) from (a) ref.~\onlinecite{Bluyssen:PLA9:482} and (b-h)
      ref.~\onlinecite{Cazzoli:CPL473:21}. Different panels show hyperfine transitions for different
      rotational bands $J'_{k_a',k_c'}\leftarrow J_{k_a,k_c}$. The measured (calculated)
      zero-crossing frequencies, in MHz, are 22235.0447 (22235.0322), 321225.6363 (321225.6311),
      380197.3303 (380197.3361), 439150.7746 (439150.7857), 443018.3358 (443018.4016), 448001.0538
      (448001.0359), 556935.9776 (556935.9849), 620700.9334 (620700.8889) for panels (a)--(h),
      respectively. }
   \label{fig2}
\end{figure*}
An overview of the calculated H$_2^{16}$O dipole absorption stick spectrum at $T=296$~K is shown in
\autoref{fig1}. The forbidden \emph{ortho-para} transitions are plotted as red circles. Despite
being, at least, 10 orders of magnitude weaker than the corresponding allowed transitions, for some
of the strongest \emph{ortho-para} transitions the predicted absorption intensities are close to the
sensitivity threshold of modern cavity ring-down spectroscopic techniques~\cite{Kassi:JCP137:234201,
   Karlovets:JQSRT247:106942, Tobias:NatComm11:1708}. All predicted \emph{ortho-para} transitions
with line intensity larger than $10^{-31}$ cm/molecule are listed in~\autoref{tab1}.
These transitions all occur in the
fundamental $\nu_2$ bending and the overtone $2\nu_1 + \nu_2$ and $\nu_1+\nu_2+\nu_3$ bands. The
off-diagonal elements of molecular-frame spin-rotation tensor $\bar{M}_{\alpha\beta, n}$, which lead
to \orthopara interaction, are highly dependent on the bending vibrational coordinate, indicating
significance of the $\nu_2$ band in \orthopara transitions. The size of the off-diagonal
spin-rotation matrix elements increases for bending angles close to \degree{180}, \ie, the linear
geometry of the molecule. This leads to an increase in the \orthopara interaction for rovibrational
energies close to the linearity barrier at $\ordsim8254~\invcm$ above the zero-point energy. The
spin-rotation coupling in these vibrationally excited states is responsible for the \orthopara
transitions. For example, the final transition state $F=3$, $J_{k_a,k_c}=4_{2,3}$ (ortho) with
energy $E=1908.016319$~\invcm is mixed with the state $F=3$, $J_{k_a,k_c}=3_{3,1}$ (para) with
energy $E=1907.450231$~\invcm. The size of the rovibrational matrix element of spin-rotation tensor,
$\mathcal{M}_{\omega,n}^{(J'l',Jl)}$ in \eqref{m_tens} for this transitions is $\pm0.95$~kHz and
$\pm6.3$~kHz ($\pm$ for $n=1,2$) for $\omega=1$ and 2, respectively. Note that following
\eqref{hsr_me} only the spin-rotation tensor with $\omega=1$ contributes to the \orthopara coupling.
Allowed transitions into these states from the ground state are quite strong, $2.07\times10^{-20}$
and $3.52\times10^{-20}$~cm/molecule, respectively. Accordingly, intensity borrowing as a result of
the spin-rotation interaction of excited states leads to non-zero intensities of the two
corresponding forbidden transitions on the order of $10^{-31}$ molecule/cm. Similarly for other of
the strongest forbidden transitions listed in \autoref{tab1}, the enhancement occurs due to
intensity borrowing effect from strongly allowed transitions with coincident near resonance between
the excited states, accompanied by a relatively large value of the spin-rotation matrix element
$\mathcal{M}_{\omega=1,n}^{(J'l',Jl)}$.

\begin{figure*}
   \includegraphics[width=0.8\linewidth]{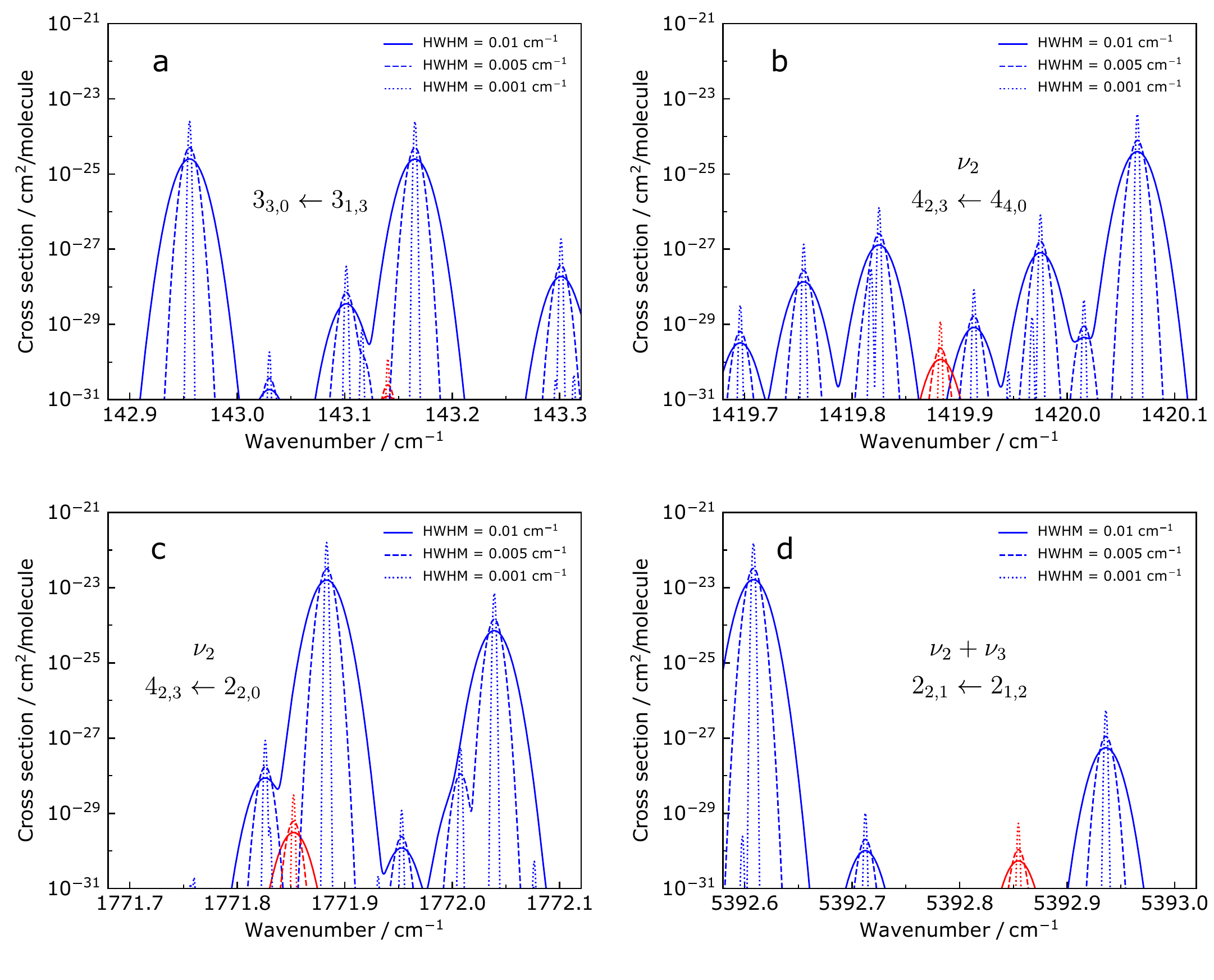}
   \caption{Absorption cross sections computed at $T=296$~K for selected rotational bands, using
      Gaussian lineshapes with HWHMs of 0.01~\invcm (solid lines), 0.005~\invcm (dashed lines),
      and 0.001~\invcm (dotted lines). The cross sections for allowed \emph{ortho-ortho} and
      \emph{para-para} transitions are plotted with blue colour lines and cross sections for
      forbidden \emph{ortho-para} transitions are plotted with red colour lines.}
   \label{fig3}
\end{figure*}

Though \emph{ortho-para} transitions are yet to be observed in \HHO, there are several spectroscopic
studies of the allowed hyperfine transitions in the pure rotational spectrum of
$\text{H}_2^{16}\text{O}$~\cite{Bluyssen:PLA9:482, Kukolich:JCP50:3751, Golubiatnikov:JMS240:251,
   Cazzoli:CPL473:21}. We used these data to validate the accuracy of our predictions.
In~\autoref{fig2} the calculated transitions (stems) are compared with the available experimental
data (dashed lines), demonstrating an excellent agreement, within 1--4 kHz, for the hyperfine
splittings. For example, the root-mean square (rms) deviation of the predicted hyperfine splittings
from experiment is 2.1~kHz in \autoref[a]{fig2}, while for the absolute line positions it is
12.3~kHz. The latter can be explained by the discrepancies in predictions of the pure rotational
transitions. The errors in predictions of the hyperfine splittings can be attributed to the level of
electronic structure theory, in particular the basis set, employed in the calculations of
spin-rotation tensor surface. The basis set convergence of the equilibrium spin-rotation constants
of \HHO was investigated elsewhere~\cite{Cazzoli:CPL473:21}. According to the results, the employed
aug-cc-pwCVTZ basis set produces an average error of 1.3~kHz with a maximum of 1.8~kHz for one of
the off-diagonal elements, when compared with the results obtained with the aug-cc-pwCV6Z basis set.
There are several predicted splittings in \autoref[d--h]{fig2} that are less than 12~kHz and were not
resolved in the experiment~\cite{Cazzoli:CPL473:21}. Indeed, by visual inspection of the Lamb-dip
spectrum plotted in Fig.~1 of ref.~\onlinecite{Cazzoli:CPL473:21}, which was provided as an example
of the experimental resolution achieved in that work, the transition profiles' full width at a half
maximum is about 13~kHz.

The sensitivity and resolution required to observe the \orthopara transitions in a prospective
experiment can be estimated from the simulated absorption spectrum, shown \autoref{fig3} for
selected wavenumber ranges with strong \orthopara transitions. Since the Doppler linewidth would be
around 0.01~\invcm at room temperature and even much higher-resolution spectroscopy was
demonstrated~\cite{Daussy:PRL83:1554}, we used simple Gaussian line profiles with half-width at
half-maximum (HWHM) fixed at 0.01, 0.005, and 0.001~\invcm and computed absorption cross sections at
$T=296$~K using \texttt{ExoCross}~\cite{Yurchenko:AA614:A131} to predict the experimental spectra.
The \orthopara transitions In \autoref[a,c]{fig3} (red) show considerable overlap with the allowed
transitions (blue) for purely rotational transitions and in the fundamental $\nu_2$ excitation band
and could only be detected with an experimental HWHM below 0.005~\invcm at an experimental
sensitivity of $10^{-30}$ and $10^{-29}$~cm$^2$/molecule, respectively. In \autoref[b,d]{fig3},
showing parts of the $\nu_2$ and $\nu_2+\nu_3$ bands, the predicted \orthopara transitions are
better separated from the allowed transitions and should already be detectable at lower resolution,
\ie, at HWHM of 0.01~\invcm, but demand a greater sensitivity of $10^{-30}$ and
$10^{-31}$~cm$^2$/molecule, respectively. Such high-sensitivity measurements of intensities on the
scale of $10^{-30}$~cm$^2$/molecule are currently within reach, for example, using continuous wave
laser cavity ring down spectroscopy~\cite{Campargue:PCCP14:802, Campargue:PRR2:023091}.

\section{Conclusions}
We developed and performed comprehensive variational calculations of the room temperature linelist
of \HHO with hyperfine resolution, including forbidden \orthopara transitions. The calculations were
based on accurate rovibrational energy levels and wavefunctions produced using the variational
approach TROVE. The nuclear hyperfine effects were modeled as spin-rotation and direct spin-spin
interactions, with the spin-rotation coupling surface calculated at a high level of the
electronic-structure theory. We found excellent agreement between the calculated transition
frequencies and available hyperfine-resolved spectroscopic data of allowed transitions.

The predicted \orthopara transitions are useful for guiding future experimental spectroscopic
studies in search of these forbidden transitions in the laboratory as well as in astrophysical
environments. Our accurate predictions of hyperfine effects complement the spectroscopic data for
water.

The variational approach we developed for computing these hyperfine effects is general. It includes
nuclear quadrupole~\cite{Yachmenev:JCP147:141101, Yachmenev:JCP151:244118}, spin-rotation, and
spin-spin interactions, and can be applied to other molecular systems without restrictions on the
number and values of nuclear spins.

\section*{Supplementary material}
The computed hyperfine-linelist data for \HHO are available at
\url{https://doi.org/10.5281/zenodo.6337130}.

\section*{Author declarations}
\subsection*{Conflict of interests}
The authors have no conflicts to disclose.

\section*{Data availability}
The computer codes used in this work are available from git repositories at
\url{https://github.com/Trovemaster/TROVE} and \url{https://github.com/CFEL-CMI/richmol}.

\section*{Acknowledgments}
We acknowledge support by Deutsches Elektronen-Synchrotron DESY, a member of the Helmholtz
Association (HGF). This work was supported by the Deutsche Forschungsgemeinschaft (DFG) through the
priority program ``Quantum Dynamics in Tailored Intense Fields'' (QUTIF, SPP1840, YA~610/1) and the
cluster of excellence ``Advanced Imaging of Matter'' (AIM, EXC~2056, ID~390715994) and through the
Maxwell computational resources operated at Deutsches Elektronen-Synchrotron DESY, Hamburg, Germany.
S.Y.\ acknowledges support from the UK Science and Technology Research Council (STFC,
No.~ST/R000476/1) and the European Research Council under the European Union’s Horizon 2020 research
and innovation programme through an Advanced Grant (883830). The authors acknowledge the use of the
Cambridge Service for Data Driven Discovery (CSD3), part of which is operated by the University of
Cambridge Research Computing on behalf of the STFC DiRAC HPC Facility (www.dirac.ac.uk). The DiRAC
component of CSD3 was funded by BEIS capital funding via STFC capital grants ST/P002307/1 and
ST/R002452/1 and STFC operations grant ST/R00689X/1. DiRAC is part of the National e-Infrastructure.
G.Y.\ gratefully acknowledges the financial support by the China Scholarship Council (CSC).

\bibliography{string,cmi}

\onecolumngrid
\listoffixmes
\end{document}